# Temperature Control of Digital Glass Forming Processes

Balark Tiwari[1], Nishan Khadka[1], Nicholas Capps[2], Andre Bos[2], Douglas Meredith[2], John Bernardin[2], Edward C. Kinzel[1], and Robert G. Landers[1]

University of Notre Dame[1]
Los Alamos National Laboratory[2]

**Abstract**
Digital Glass Forming (DGF) is a new manufacturing process for low-batch glass fabrication. The work zone temperature in DGF processes must be maintained in the glass's working range to ensure good fabrication. If the temperature is too low, the filament will not wet to the substrate or previously deposited material and, if the temperature is too high, the filament may disengage from the substrate or previously deposited material, or it may partially vaporize. In this work, a real-time temperature control system capable of synchronizing process parameter, thermal camera, and visual camera data for the DGF process is introduced. A process parameter map for a scan velocity of 0.5 mm/s is constructed, as is a data-driven dynamic temperature process model. A digital controller is designed to regulate the work zone temperature. The temperature controller is a closed loop tracking controller that adjusts the commanded laser power to regulate the measured temperature. Two sets of experiments are conducted to analyze the controller performance. In the first set of experiments, single tracks on a substrate are fabricated with constant laser power and with the closed loop temperature controller. It is seen that the closed loop controller is able to extend the process parameter map into regions where using a constant laser power will result in a failed build. In the second set of experiments, walls are fabricated. Using constant laser power results in a failed build (i.e., material vaporization at the corners and the filament prematurely detaching from the substrate) as the temperature process dynamics change with layer and at the corners. The closed loop controller successfully fabricated the wall without vaporization at the corners and premature filament detachment as the controller adjusts the laser power to account for the changing temperature process dynamics.

## 1. Introduction

Glass fabrication methods can be broadly categorized into high- and low-batch. Traditional high-batch industrial methods, such as blow molding (for bottles and lamps), drawing (for tubes and fibers), and floating (for flat sheets), are efficient but limited to simple geometries and often requiring extensive post-processing (e.g., grinding, polishing, drilling, coating, etching) for improved surface accuracy [1]. In contrast, low-batch artisanal methods, e.g., scientific glassblowing for complex apparatuses, leverage skilled manipulation of viscous glass with gravity or hand tools, yet suffer from large variability, low repeatability, and production rates typically below 10 pieces per day. The complex material properties of glass have hindered robotic or autonomous solutions that combine the precision and speed of CNC technologies with the ability to create intricate geometries. Digital Glass Forming (DGF), incorporating additive, subtractive, and reshaping techniques, offers the capability to address these challenges by enabling the precise fabrication of parts with complex geometries.

Precise temperature control is critical in DGF processes due to the viscoelastic behavior of glass, where viscosity decreases exponentially with temperature [2,3], from $\sim 10^{13}$ Pa·s at the strain point ($\sim 500$ °C) to $10^4$–$10^7$ Pa·s at working temperatures ($\sim 900$–$1100$ °C) for soda lime glass [4].

At working temperatures, controlled flow and substrate wetting are facilitated; however, there is an increased risk of defects like cracking, bubble entrapment, and warping if thermal gradients are not properly managed [5,6]. This sensitivity requires real-time adjustments to process parameters and pathing to counter disturbances like filament deflection and complex, fast thermal and spatial thermal gradients. High-precision DGF requires multi-modal sensor fusion, combining force feedback (for pressure), visual sensing (for geometry and layer height), spatial thermal imaging (for work zone localization), and confocal scanning (for high resolution morphology) to enable real time feedback process control. Prior work in this area includes limited closed loop approaches. Chen et al. [7] and Peters et al. [8] used feedback controllers with pyrometer feedback to stabilize the work zone temperature; however, they faced the challenge of keeping the single-point sensor continuously aimed at the work zone. Most studies rely on pseudo-control, such as manual laser adjustments [9] or design-of-experiments [8,10] that require extensive experimental trials to search the process parameter space and lack the adaptability needed to efficiently reject disturbances such as filament diameter variations and deflections. To address these limitations, this work utilizes spatial thermal imaging and control-oriented modeling to develop temperature control systems that enable robust deposition in DGF processes and substantially reduce process development time.

Chen et al. [7] explored temperature control in glass additive manufacturing of quartz via direct energy deposition with coaxial wire feeding. They implemented a PID closed loop control system that adjusted laser power in real-time using pyrometer feedback to stabilize the molten pool temperature, thereby reducing defects such as bubbles and pores while improving surface microstructure and molding quality over tests with constant laser powers. However, the use of a pyrometer created susceptibility to emissivity variations and laser spot misalignment, and introduced calibration challenges, which limited the precision they were able to achieve. Zhang et al. [11] employed selective laser melting to fabricate multi-material components combining soda-lime glass and metal, identifying cracking issues attributed to unmanaged thermal stresses in the soda lime glass, and created a laser power selection approach to mitigate these defects. However, their approach cannot dynamically adjust process parameters, which limits the precision and repeatability needed to effectively address thermal stress challenges. Capps et al. [6] explored deposition of hollow borosilicate tubes for microfluidic and photonic applications, optimizing laser power, tube feed rate, and pneumatic pressure through design of experiments to achieve precise deposition and prevent tube collapse. The study emphasized the role of the temperature distribution profile in the laser-heated zone for high-quality deposition, employing an open loop control approach for these process parameters. Liu et al. [12] utilized a filament-based 3D printing approach with CO2-laser heating to fabricate multi-layer fused silica-glass structures, relying on an open loop control strategy where printing speed and incident laser power were experimentally optimized to enhance bonding width and layer integrity, achieving structures with over 100 layers. This work showed that the incorrect choice of process parameters such as printing speed, filament feed rate, and incident laser power may introduce challenges in the fabrication process including excessive vaporization and poor geometry control, issues that could potentially be mitigated through the incorporation of feedback control. Lin et al. [13] utilized a six-axis KUKA KR60 robotic arm for glass 3D printing of borosilicate glass, noting transparency and warping issues with multilayer structures due to uncontrolled thermal variations. Luo et al. [14] developed a filament-fed additive manufacturing technique using a CO2 laser to melt soda-lime glass filaments and employed a heated build platform and open loop control with manual laser power adjustments based on a thermal model to optimize transparency and reduce defects like bubbles. Peters et al. [7] built a glass additive manufacturing system using a CO2 laser and pyrometer to sense the molten pool, incorporating a heated build platform and motion system to create multi-layer structures like thin-walled stars and free-standing springs, while addressing bubble formation and melt pool displacement. They employed a closed loop temperature controller using loop shaping to maintain stable melt pool conditions and a path planning algorithm to ensure consistent filament feed direction, enhancing optical and dimensional accuracy across varying build speeds.

This paper introduces a novel experimental system for DGF temperature control by integrating deterministic real-time control and spatial thermal imaging. Unlike prior work relying on manual process parameter tuning, open loop strategies, or single-point pyrometer measurements [7,8,14], contributions of this paper are: a real-time control system built on an RT-Linux platform, enabling the synchronized integration of heterogeneous sensors and actuators with different interfaces (e.g., Ethernet/IP, GigE Vision, Digital and Analog IO); in-situ thermal monitoring using a thermal camera, including spatial data processing for superior process stability compared to systems using pyrometers, and control-oriented modeling of the work zone temperature in volumetrically-heated DGF, elucidating thermal physics and facilitating real-time control design. Another contribution is the use of closed loop temperature control to expand the process parameter region, eliminating the need for extensive design-of-experiments and enabling stable fabrication of complex multi-layer glass structures. The paper is organized as follows: Section 2 describes the experimental setup, including hardware components (optics, filament feeder, motion stages, and sensors), in-situ thermal measurement challenges, and volumetric laser beam modeling. Section 3 presents thermal process modeling, covering process parameter mapping and control-oriented model identification and validation. Section 4 details the controller design and Section 5 presents experimental studies, demonstrating controller effectiveness in track fabrication outside of established process parameter maps and for multi-layer structures. Finally, Section 6 provides a summary of the paper and draws conclusions.

## 2. Experimental Setup

The DGF system utilized in the experimental studies conducted in this paper (Figure 1) consists of a custom optical assembly, a high-power fiber laser, a filament feeder, and a motion system for precise positioning. The Automation1 iXC4 and PRO115SLE motion system (Aerotech) enables positioning with 1 μm repeatability and 16 μm accuracy. The custom optical assembly first expands the incident laser beam and subsequently employs an off-axis parabolic mirror (Thorlabs) to focus it into a converging conical beam. This configuration generates an extended three-dimensional focal region, referred to as the volumetric cone, along the beam propagation direction, thereby enabling volumetric heating of the glass filament via bulk absorption, as opposed to the predominantly surface-limited heating characteristic of conventional focused-spot approaches. Laser input is provided by a 500 W single-mode ytterbium fiber laser YLR-500 (IPG Photonics) with a wavelength of 1070 nm. A custom filament feeder accommodates glass stock filaments ranging from 0.5 to 2.0 mm in diameter. Driven by an Aerotech BM75 motor, the feeder allows variable feed rates and precise filament positioning. A PM Plus heater controller with 200 W firerod cartridge heaters (Watlow) and a custom-machined heater plate maintain the heatbed at 550°C during glass deposition to anneal the glass while it is being deposited. An Oryx ORX-10G-310S9C color 10GigE camera (FLIR), equipped with a 0.28×28.7 mm PlatinumTL telecentric lens (Edmund Optics), provides high-speed optical imaging of the work zone at up to 417 fps (with reduced sensor area) and a 12.4 μm effective pixel size. An Everest K2 chromatic confocal sensor (Acuity) enables top-down morphology scanning of glass objects with ±0.15 μm accuracy, a 2 mm measuring range, and up to ±42° slope angle capability. Soda-lime glass stringers (Bullseye Glass) with nominal diameters of 1.0 mm and turquoise blue color were used as the feed filament, while clear soda-lime, glass slides (Ted Pella) were used as substrates. The chosen glass stringers exhibit strong absorption in the near-infrared spectrum and an optical penetration depth of approximately 2.13 mm [15], facilitating near uniform heating across the filament.

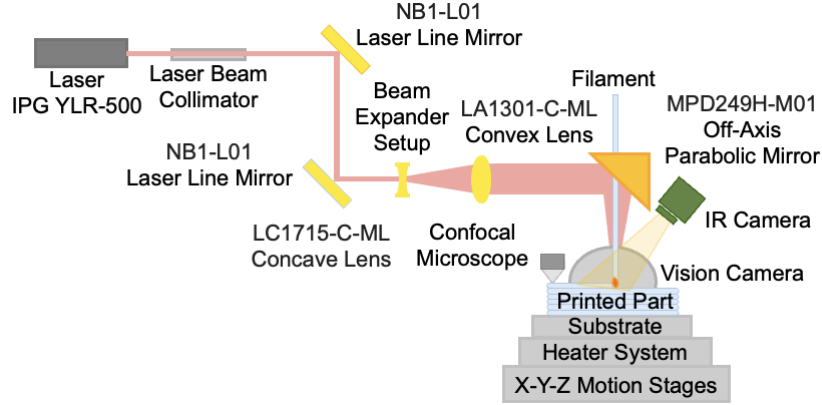

**Figure 1:** Schematic of laser-heated glass printing setup.

## 2.1. In-Situ Thermal Measurement

The process monitoring system employs a PI 640i thermal camera (Optris) with a frame rate up to 125 Hz, 148 µm/pixel spatial resolution, and temperature measurement capabilities up to 1500 °C. This setup enables detailed observation of spatial thermal distributions and temporal trajectories during heating and cooling, including the thermal history of previously deposited layers. The camera incorporates an uncooled microbolometer infrared detector operating in the long-wave infrared range (8–14 µm), equipped with a 15° × 11° optic (41.5 mm focal length). It provides high-resolution thermal imaging with a noise-equivalent temperature difference of 40 mK. Non-uniformity correction via an internal flag mitigates sensor drift, though this process requires approximately 128 ms (plus time for mechanical flag operation), potentially impacting real-time measurements. This is a common limitation of microbolometer-based thermal imaging systems. In contrast to pyrometers, which function as single-point sensors and lack dynamic region-of-interest (ROI) adjustment [8,16], the thermal camera's 640 × 480-pixel array covers a 94.7 × 70.4 mm field of view. This encompasses the filament feed, work zone, and previously deposited glass, enabling spatial temperature mapping and identification of the hottest pixel clusters to locate the work zone accurately. For a 1 mm diameter filament, averaging the highest 200 hottest pixels was empirically determined to represent the work zone. Advantages of thermal imaging include the possibility of voxel-by-voxel temperature profiling, which facilitates measurement of heating and cooling rates, critical for mitigating structural cracks through optimized thermal profiles, and estimation of bulk temperatures from surface readings. However, the disadvantage of thermal cameras is their higher computational demands for data communication and processing. Pyrometers, while simpler and faster for point measurements, cannot account for spatial variability. The emissivity of soda-lime glass varies with temperature, ranging from 0.6–0.8 at room temperature to 0.85–0.95 at elevated temperatures (~900°C), highlighting the need for further research to determine precise values [17]. In this study, an average emissivity of 0.72 was used for thermal camera measurements [16].

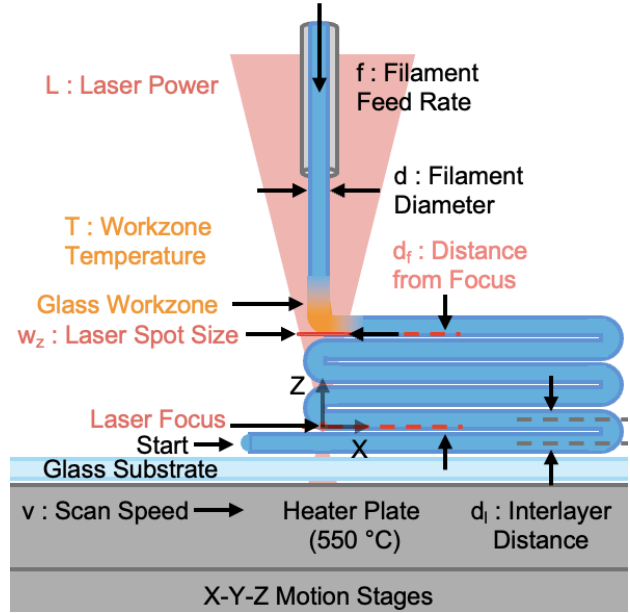

**Figure 2:** Schematic illustration of DGF process. Illustration shows laser power ($L$), filament feed rate ($f$), filament diameter ($d$), work zone temperature ($T$), distance from focus ($d_f$), interlayer distance ($d_l$), and scan speed ($v$). Setup shows glass work zone, fabricated 6-layer wall, laser spot size, laser focus, glass substrate, and heater plate.

To compare pyrometry to thermal imagery, square chimney structures, 20 mm × 20 mm with four layers, are fabricated with $v$ = 0.5 mm/s, $d_f$ = 4 mm, and $d_l$ = 1.5 mm. The deposition strategy involved continuous track deposition per layer along a fixed rectangular path: the XY stages traversed 20 mm (−X), 20 mm (−Y), 20 mm (+X), and 20 mm (+Y), returning to the starting corner. After each layer the Z-stage was lowered by 0.8 mm (-Z). This counterclockwise path was repeated for all layers. Deposition began after the filament reached $T_r$ = 900 °C under closed loop control (described below) and the first layer was deposited directly onto the preheated glass substrate. Following the final layer, motion ceased, feeding stopped, and filament retraction was performed at $f$ = -2.0 mm/s. To emulate a pyrometer, the Lumasense IMPAC pyrometer, utilized in [8,18] and having a spot diameter of 0.9 mm [7,18], was employed. A circular ROI of this diameter was defined on the thermal camera frames and the temperature was computed as the average of the pixels in the ROI (Figure 3 and Figure 4). Note the ROI is fixed in the thermal camera frame.

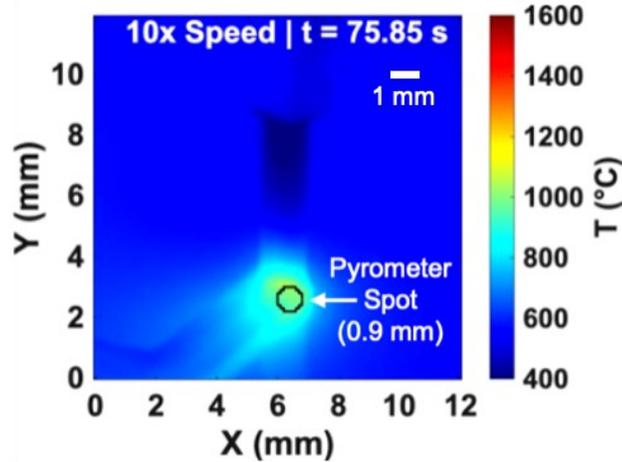

**Figure 3:** Thermal camera view of digital glass forming process with simulated pyrometer spot.

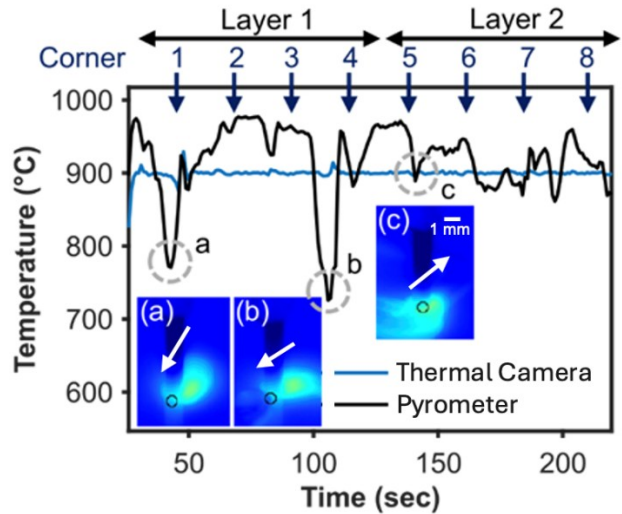

**Figure 4:** Observed temperature values by pyrometer simulation and thermal camera during temperature-controlled chimney deposition. Insets show thermal camera images at various times. Arrows in the insets indicate deposition direction.

In the first layer (Figure 3), manual alignment placed the pyrometer spot near the expected work zone center. The limitation of pyrometer sensing becomes evident during directional changes in deposition where the filament substantially deflects (Figure 4). At $t = 41$ and 105 sec, the filament momentarily exits the pyrometer spot, causing an apparent temperature decrease in excess of 100 °C. In contrast, the thermal camera consistently tracks the work zone even when it moves in the thermal camera frame. Representative thermal images (insets in Figure 4) illustrate three scenarios: (a) and (b) filament outside the pyrometer spot (pyrometer fails) and (c) ideal alignment where both sensors agree. The simulated pyrometer readings occasionally exceed those reported by the thermal camera (Figure 4). This discrepancy arises because the pyrometer averages the radiance over a fixed circular spot of only 0.9 mm diameter (approximately 30 pixels), which is smaller than the 200 pixels averaged by the thermal camera. Consequently, the pyrometer spot can capture localized regions of higher temperature concentration. Thermal camera monitoring provides a more robust, representative measurement of the work zone temperature in filament-fed glass deposition, particularly when the filament deviates from its nominal position, where a fixed-spot pyrometer can temporarily lose the hottest region entirely (insets a and b in Figure 4).

## 2.2. Laser Beam Modeling and Filament Interaction

The initial collimated beam diameter from the fiber is 8.64 mm. To improve the focused beam quality and minimize the Rayleigh length, the collimated beam diameter was increased to 43.17 mm before entering the OAP. Beam propagation through the optical chain was modeled using the ABCD matrix formalism under paraxial and thin-lens approximations [18]. The beam travels through several free-space segments in air and passes through a plano-concave lens with a focal length of –50 mm and a diameter of 25.4 mm, a plano-convex focusing lens with a focal length of 250 mm and a diameter of 50.8 mm, and finally an off-axis parabolic mirror with a reflected focal length of 101.6 mm. Numerical propagation of the complex beam parameter through this sequence yields a tightly focused Gaussian spot at the laser focus with a diameter of 3.2 μm. The resulting beam radius is

$$w_z(z) = w_0 \sqrt{1 + \left(\frac{z}{z_R}\right)^2} \tag{1}$$

where the Rayleigh range is $z_R = 7.544$ μm and the beam waist at the laser focus is $w_0 = 1.603$ μm. These parameters were calculated analytically by propagating the complex beam parameter (i.e., $q$-parameter). The on-axis intensity distribution incident on the filament and substrate follows the Gaussian intensity distribution

$$I(r,z) = \frac{2L}{\pi w_z^2(z)} \exp\left(-\frac{2r^2}{w_z^2(z)}\right) \tag{2}$$

where $L$ is the commanded laser power (W), $r$ is the radial distance from the center of the laser beam, and $z$ is the vertical distance from the focal point (see Figure 2).

Normalized radial profiles of the power absorbed by the filament at various distances from the focus are shown in Figure 5 and an image of the filament overlayed with the laser intensity field is shown in Figure 6. The laser spot size has large implications on the DGF process. For small spot sizes, i.e., small values of $d_f$, the beam is tightly focused, producing high peak intensities (> $10^3$ W/mm²) and large radial gradients (Figure 5). Even sub-millimeter lateral misalignments of the 1 mm-diameter glass filament causes drastic losses in the absorbed power (Figure 5), thus, the melt pool is highly sensitive to lateral filament and hypodermic tube displacements. However, a small spot size prevents the heating of material outside of the work zone. For larger spot sizes, i.e., large values of $d_f$, the smaller peak intensities and radial gradients significantly reduce sensitivity to lateral filament and hypodermic tube displacements However, previously deposited glass around the work zone will be heated. Empirically, a spot size range of $0.85 \leq w_0 \leq 1.91$ mm (i.e., $4 \leq d_f \leq 9$ mm) has been found to provide good track deposition for the studies conducted in this paper.

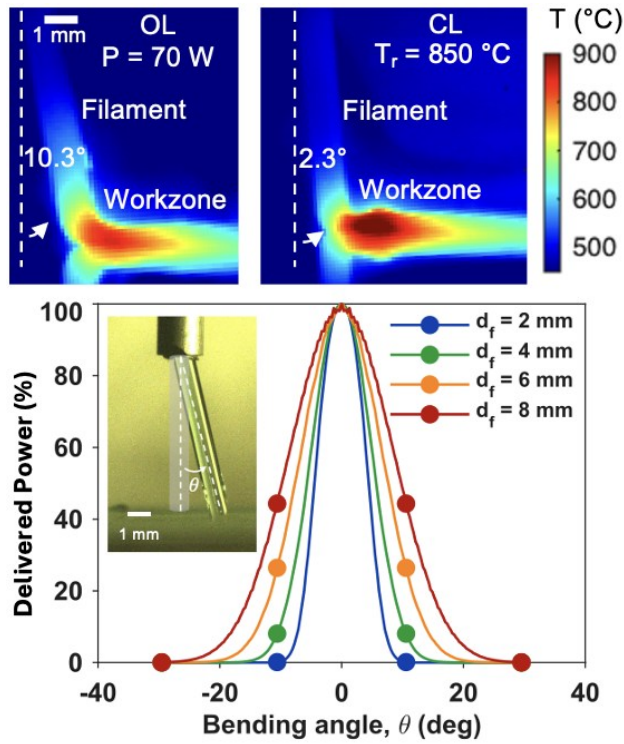

**Figure 5:** Normalized percent power absorbed by the filament relative to on-axis maximum inside 1 mm diameter circle versus bending angle $\theta$ for $d_f$ = 2, 4, 6, and 8 mm. Top insets show thermal images with bending angle $\theta$ annotated: (left) open loop (OL) case just before print failure by filament fracture and (right) closed loop (CL) case at same time stamp with successful print. Experiments are described below.

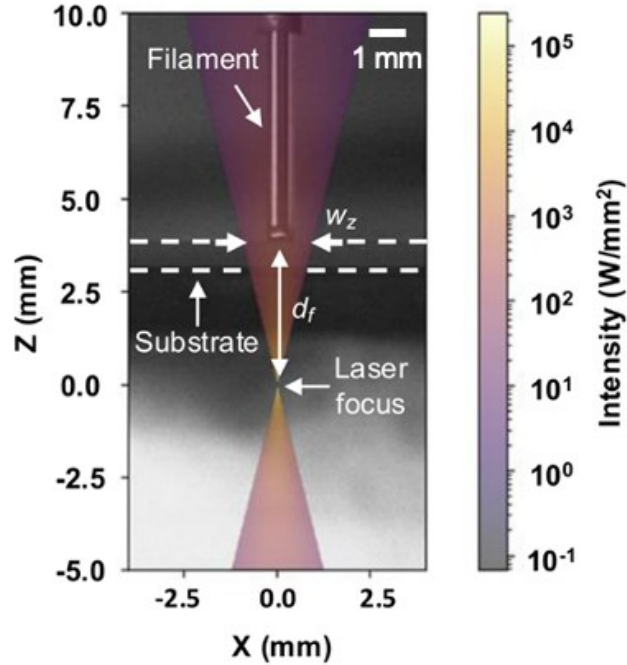

**Figure 6:** Spatial laser intensity distribution in x–z plane overlayed on DGF system image (logarithmic color scale). Intensity range is restricted to $10^{-1} - 10^5$ W/mm² to highlight distribution within practical glass processing regime. Note peak intensity at laser focus is ~$10^8$ W/mm².

## 3. Thermal Process Modeling

### 3.1. Process Parameter Mapping

A process parameter map was generated for $20 \leq L \leq 60$ W using increments of 10 W and $4 \leq d_f \leq 9$ mm using increments of 1 mm for $v = 0.5$ mm/s by fabricating tracks on a substrate. For each process parameter combination, the maximum temperature is shown with the heatcolor corresponding to the maximum temperature. The maximum temperature increases with increasing laser power and decreasing distance from the focus. All sets of process parameters denoted with an x resulted in failed depositions. The depositions with ($L = 20$ W and $d_f = 8$ mm) and ($L = 10$ W and $d_f = 3$ mm) had work zones that were too cold to wet to the substrate. For the first set of process parameters, the filament remained rigid, continually slipping in the feeder and dragged along the substrate. For the second set of process parameters, the filament became warm enough to bend, but was not hot enough to fuse to the substrate, curling up without wetting to the substrate. The depositions with ($L = 60$ W and $d_f = 5$) and ($L = 70$ W and $d_f = 7$ mm) were too hot to create proper tracks. For the first set of process parameters, the work zone disengaged from the substrate during deposition due to surface tension while, for the second set of process parameters, material was partially vaporized during deposition.

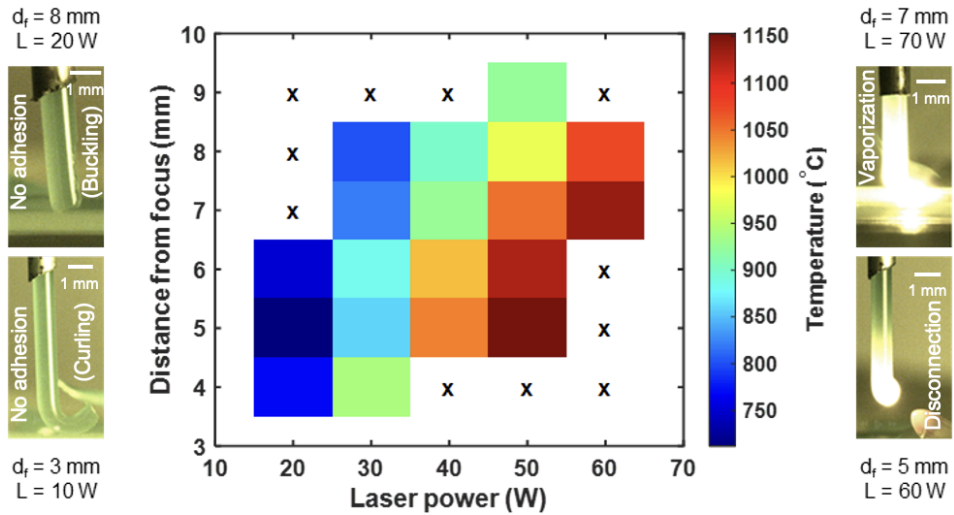

**Figure 7:** Process parameter map in $L$-$d_f$ plane with $v = 0.5$ mm/s showing maximum temperatures. Images illustrate various failure modes.

### 3.2. Data-Driven Modeling and Model Validation

A lumped-parameter analytical model for the work zone temperature in filament-fed glass additive manufacturing, balancing laser energy input with conduction, convection, advection, and substrate losses, was previously derived by Luo et al. [19]. Motivated by this model structure, a first-order linear dynamic model relating commanded laser power to measured temperature is adopted in this study. However, due to uncertainties in thermophysical parameters (e.g., effective conductivity, convection coefficients), a data-driven system identification approach is employed to construct the dynamic model.

    The thermal DGF process exhibits significant, naturally occurring variations during deposition. One source is the temperature dependence of model coefficients. For example, the specific heat capacity of silicate glasses changes with temperature due to increased vibrational energy near the glass transition temperature [20]. The thermal conductivity and diffusivity of glass decrease with temperature near the glass transition temperature due to increased phonon scattering induced by structural disorder [2]. Additionally, glass emissivity varies nonlinearly with temperature, rising sharply for temperatures above 800 °C due to increased infrared emission from dynamical disorder in the silicate network [2]. Radiation losses, governed by the Stefan-Boltzmann law, becomes increasingly dominant at high temperatures, creating deviations from linear behavior. Fluctuations in filament diameter, arising from manufacturing tolerances, introduce stochastic variations to the area over which heat conducts and convection occurs. Deviations in process parameters, such as scan velocity and filament feed rate, from the nominal values used to establish the process parameter map perturb the expected heat transfer. Boundary condition variations, including substrate temperature gradients across the heater plate and deposition near substrate edges or on previously deposited material, significantly alter spatial thermal distributions. Furthermore, the relative motion between the filament and laser beam profile, coupled with the glass's viscoelastic properties that drive dynamic reflow, modifies the work zone geometry, as evidenced by track morphology variations. These factors underscore the necessity for a simplified, control-oriented model. While detailed simulations capturing the complex thermal-mechanical

evolution of the glass during the deposition process can be constructed, they are computationally demanding and offer limited applicability to real-time control systems, which require rapidly executable models for in-situ real time control and optimization.

Data-driven techniques are now used to construct a temperature model of the DGF process relating the average of the maximum 200 hottest thermal camera pixels to the commanded laser power. Assuming constant scan velocity and filament feed rate, the model structure for the temperature process dynamics is

$$a\Delta\dot{T}(t) + \Delta T(t) = b\Delta L(t) \tag{3}$$

where $\Delta T(t) = T(t) - T_n$ (°C), $\Delta L(t) = L(t) - L_n$ (W), $T_n$ is the nominal temperature (°C), $L_n$ is the nominal laser power (W), $a$ is the time constant (sec), and $b$ is the gain (°C/W). Model parameters were determined via system identification techniques using experiments designed to span the process parameter range used to deposit tracks on substrates. During the identification process, $v = f = 0.5$ mm/s, $d = 1.0$ mm, the heat bed was maintained at 550 °C, $30 \leq L \leq 60$ W, and $d_f = 7.0$ mm, yielding an approximate laser spot diameter of 3.5 mm. Each experiment deposited a 60 mm track, and the initial and final 10 mm of the track were excluded from the analysis to focus on steady deposition dynamics. System identification laser power trajectories, i.e., (a) a Pseudo Random Binary Signal (PRBS) with low and high laser power values of 30 and 60 W, respectively, and spanning a frequency range of 0–5 Hz, (b) a 0–0.2 Hz chirp signal, and (c) a 31.4 rad/s sinewave signal were generated, applied in track fabrication, and analyzed using Matlab's System Identification Toolbox. Figure 8 shows the results of each test.

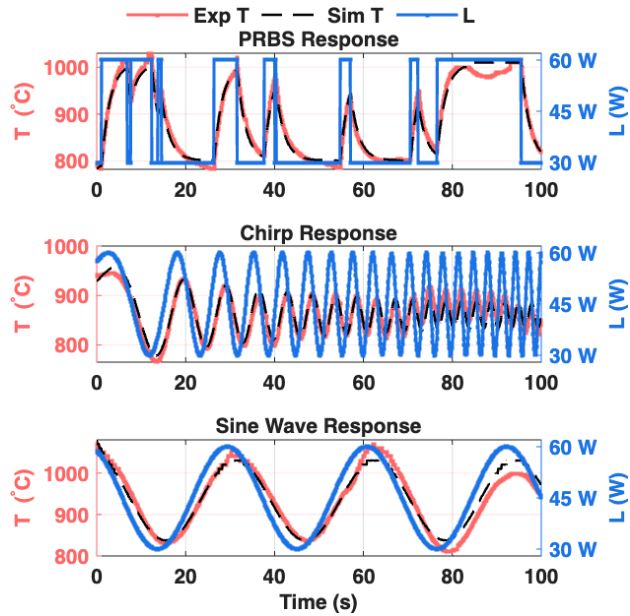

**Figure 8:** Measured and simulated temperature responses for system identification experiments (a) PRBS between 30 and 60 W, (b) chirp (0–0.2 Hz) data using $f_s = 10$ Hz, and (c) sinewave (31.4 s period), 60 mm long track, $v = 0.5$ mm/s, $d_f = 7.0$ mm, and $d = 1.0$ mm.

The experiment with the PRBS laser power signal is used to construct the temperature model around the nominal point $L_n = 42.6$ W and $T_n = 888$ °C. This operating point was chosen because it lies at the center of the only experimentally identified regime in which the DGF process exhibits stable, continuous, and approximately linear deposition behavior, characterized by a fully softened viscoelastic filament that maintains coherent attachment to the substrate via a steady work zone without detachment from the substrate or excessive vaporization. Substantial deviations from this point induce abrupt qualitative changes in process physics: laser powers well above $L_n$ elevate the work zone temperature beyond the vaporization threshold causing filament detachment, whereas laser powers significantly below $L_n$ force cold filament into the substrate, resulting in mechanical scraping, buckling, or fracture. Consequently, the identified first-order linear model is valid within this stable window. The identified transfer function is

$$\frac{\Delta T(s)}{\Delta L(s)} = \frac{3.69}{0.53s+1} \tag{4}$$

where $s$ is the Laplace operator. Validation against the chirp and sinewave signals showed > 70% accuracy. There are deviations in the trials in Figure 8 due to (i) filament diameter variations arising from manufacturing tolerances (0.85–1.5 mm for a nominally 1 mm hand-pulled glass stringer), which strongly perturb glass mass flow rate and effective work zone thermal capacitance, (ii) unmodeled convective currents and small laser–filament misalignments, and (iii) temperature-dependent thermophysical properties and nonlinear radiative losses. These effects are difficult to capture in a low-order model amenable to real-time control and, thus, are intentionally omitted.

The controller will be implemented on a digital control platform; therefore, to design the controller the continuous-time transfer function was converted to a discrete-time transfer function via a Zero-Order Hold with sample period of $\Delta t = 0.1$ sec

$$\frac{\Delta T(z)}{\Delta L(z)} = \frac{b(z)}{a(z)} = \frac{0.6304}{z-0.8296} \tag{5}$$

where $z$ is the forward shift operator.

### 4. Controller Design

The closed loop temperature controller employs real-time temperature feedback via the thermal camera to regulate the work zone temperature at a constant value. The thermal camera captures the spatial temperature distribution during deposition and transmits the average of the hottest 200 pixels, representing a 4.3 mm² area encompassing the work zone, to the controller as the measured temperature.

The closed loop temperature controller block diagram is given in Figure 9. For $v = 0.5$ mm/s, $d_f = 7.0$ mm, and $d = 1.0$ mm, the temperature process transfer function is given by Eq (5). The control law is

$$\Delta L(z) = \frac{a(z)}{b(z)} \Delta T_r(z) - \frac{g(z)}{v(z)b(z)} \Delta E(z) \tag{6}$$

where $\Delta T_r(z) = T_r(z) - T_n$ is the incremental reference temperature (°C), $T_r$ is the reference temperature (°C), $\Delta E(z) = \Delta T_r(z) - \Delta T(z)$, $\Delta E$ is the incremental temperature error (°C), and $\Delta T$ is the incremental work zone temperature (°C). Note $\Delta E(z) = E(z)$, where the temperature error (°C) is $E(z) = T_r(z) - T(z)$. The polynomial $g(z)$ shapes the error dynamics and is

$$g(z) = v(z)a(z) - \alpha(z) \tag{7}$$

where $\alpha(z)$ is the desired closed loop characteristic polynomial. To robustly track constant references and reject constant and slowly varying disturbances, the Internal Model Principle [21] is utilized and $v(z) = z - 1$. From experimental tuning, the desired closed loop time constants are 0.1 and 0.5356 sec, yielding $\alpha(z) = z^2 - 1.197z + 0.3052$ for a sample period of 0.1 sec. It should be noted that the slower time constant is approximately five times the sample period.

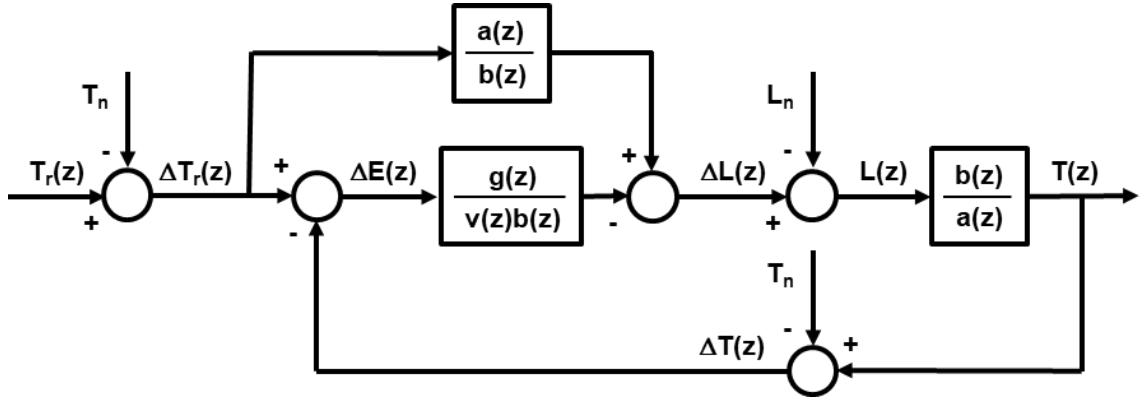

**Figure 9:** Closed loop temperature controller block diagram.

## 5. Experimental Studies

The temperature controller is implemented to fabricate straight tracks and walls. The substrate is soda-lime glass with a thickness of 1.1 mm, and the soda lime filament diameter is nominally $d = 1.0$ mm. The heatbed is set to 550°C to keep the substrate temperature close to the annealing temperature of soda lime glass.

### 5.1. Track Fabrication outside Stable Process Parameter Region

In the first set of experiments, two tracks are deposited for $f = v = 0.5$ mm/s and $d_f = 10$ mm. The first track uses $L = 70$ W and the second track uses the closed loop controller with $T_r = 800$ °C. Note that the constant-power parameter set is outside the viable process zone identified in Figure 7. As such, this deposition exhibited a failure mode where the filament periodically wets to the substrate, highlighted in Figure 10 below. In image (a), there is stable deposition. In image (b), the filament moved laterally to the left and the work zone has become cold. Severe bending of the filament has occurred in image (c); however, the filament is workable enough to bend into a curve and a work zone has again been established in image (d). By image (e), stable deposition has again been established. The closed loop controller with $T_r = 800$ °C is able to prevent this failure mode from occurring during deposition, allowing for the continuous deposition of a consistent morphology outside of the viable open-loop process window, as shown in Figure 11.

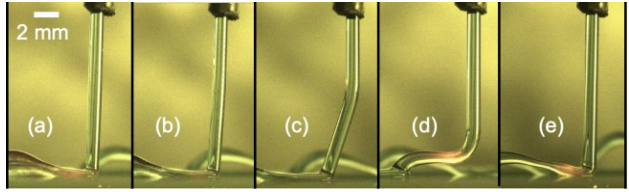

**Figure 10:** Sequential images of periodic filament instability during open loop deposition ($L = 70$ W, $f = v = 0.5$ mm/s $d_f = 10$ mm).

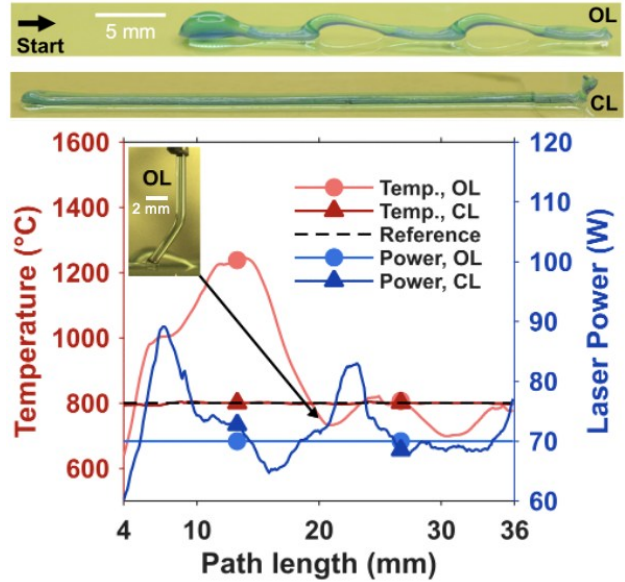

**Figure 11:** Spatial temperature and laser power trajectories for open loop ($L = 70$ W) and closed loop ($T_r = 800$ °C) depositions for $f = v = 0.5$ mm/s $d_f = 10$ mm. Images show final parts fabricated with $L = 70$ W (OL) and $T_r = 800$ °C (CL), and in-process OL fabrication.

A second set of experiments was conducted for $f = v = 0.5$ mm/s and $d_f = 3$ mm where $L = 10$, 20 and 30 W were used to fabricate three tracks. The results are shown in Figure 12. The fabrication conducted with $L = 10$ W failed because the filament was too cold to wet to the substrate, creating only a scratch on the substrate. The fabrication conducted with $L = 30$ W failed because the filament became too hot and disengaged from the substrate. For $L = 20$ W, the filament wetted to the substrate during the first half of the deposition, but disengaged from the substrate during the second half of the deposition. The closed loop temperature controller was then utilized with $T_r = 850$ °C and a stable deposition was produced. During this track fabrication, the average laser power was 20 W. At $d_f = 3$ mm, small changes in laser power (on the order of ±10 W) produce shifts in filament temperature of nearly 200 °C. Since the laser spot size at $d_f = 3$ mm and the filament diameter are both approximately 1 mm, power delivery is tightly coupled to the spatial alignment of the filament such that even slight lateral filament and hypodermic tube displacements cause large changes in the power applied to the filament and deposition can become unstable. The closed loop temperature controller efficiently expanded the process parameter map by identifying a region in which a varying laser power can be used to fabricate a track.

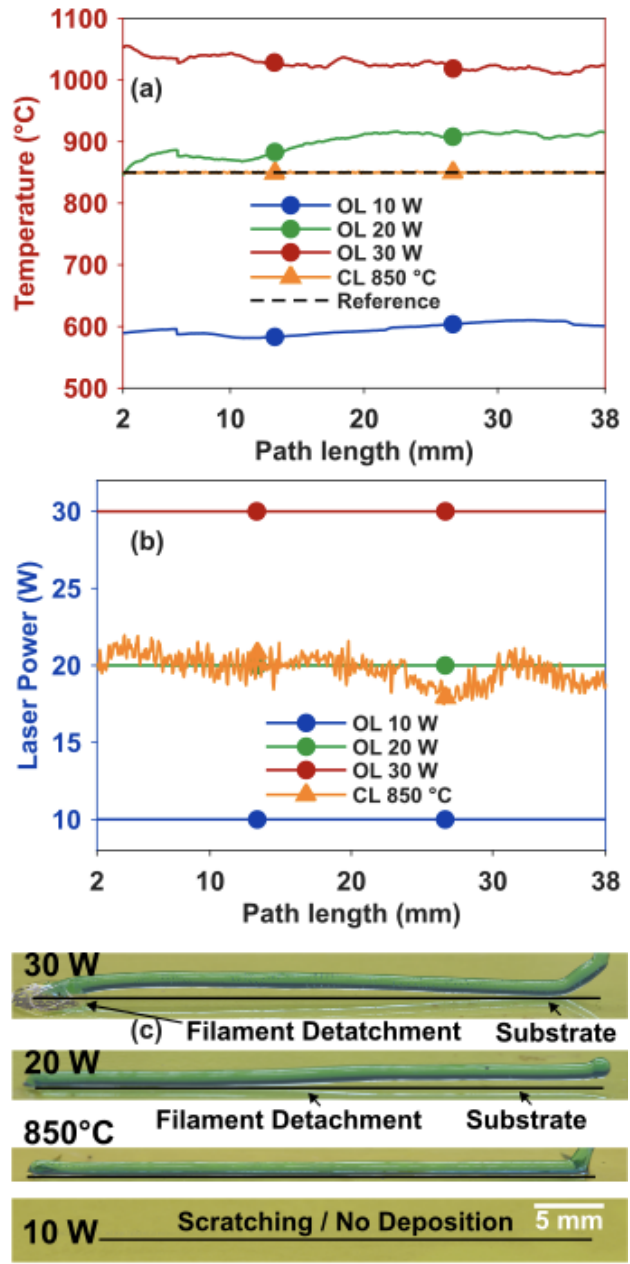

**Figure 12:** Spatial temperature (a) and laser power (b) trajectories for uncontrolled ($L$ = 10, 20, and 30 W) and controlled ($T_r$ = 850 °C) depositions for $f = v = 0.5$ mm/s and $d_f = 3$ mm. Images (c) show final tracks. Note track for $L$ = 30 W part was glued back to substrate.

### 5.2. Fabrication of Walls with Closed Loop Controller

Walls, 20 mm long having 16 layers, are fabricated for $f = v = 0.5$ mm/s, $d_l = 0.6$ mm, and $d_f = 3$ mm with constant laser power and with closed loop temperature control. The process is shown in Figure 2. The first track has a 5 mm lead-in section so that the shape of the initial deposition and the start procedure did not influence the wall fabrication. The results are shown in Figure 13. For the experiment with $L$ = 40 W, the first few layers were deposited successfully as informed by the process parameter map. However, as the build height increases the ability to conduct heat from the work zone decreases, causing the work zone temperature to increase. Also, after exiting the corners, glass is deposited on hot material, unlike deposition in the middle of the layer where the temperature of the previously deposited glass has decreased below the working temperature and reached its steady state. Thus, the input energy to the material deposited at the corners increases, dramatically raising the work zone temperature. Further, the excess heat accumulation causes pronounced rounding of the corners. This displaced material spreads laterally causing an unintended increase in the track width at the corners. As seen in Figure 13, at the ninth corner the work zone temperature increased to the point the filament detached from the part and the print failed. When the closed loop controller ($T_r$ = 940 °C) was implemented, successful deposition of all sixteen layers was achieved without filament detachment or vaporization at the corners (Figure 13). Further, excessive heat at the corners was mitigated, leading to straighter wall edges and less displaced material. As the build height increases, less laser power is required to maintain the reference temperature. As a result, the closed loop controller automatically reduces the commanded laser power from approximately 40 W during the first layer to an average of approximately 23 W during the second and subsequent layers. The closed loop controller is continuously adjusting the laser power to compensate for small lateral filament and hypodermic tube displacements, filament diameter variations, and local irregularities created from prior layers. In contrast, constant laser power deposition is highly sensitive to such disturbances: lateral filament and hypodermic tube displacements, coupled with surface roughness from previous layers, produce highly irregular power absorption and pronounced intra-layer temperature oscillations that ultimately precipitate build failure. The closed loop temperature controller not only prevents filament detachment from the substrate and vaporization at the corners due to large changes in temperature, it also yields more repeatable work zone temperature trajectories in subsequent layers.

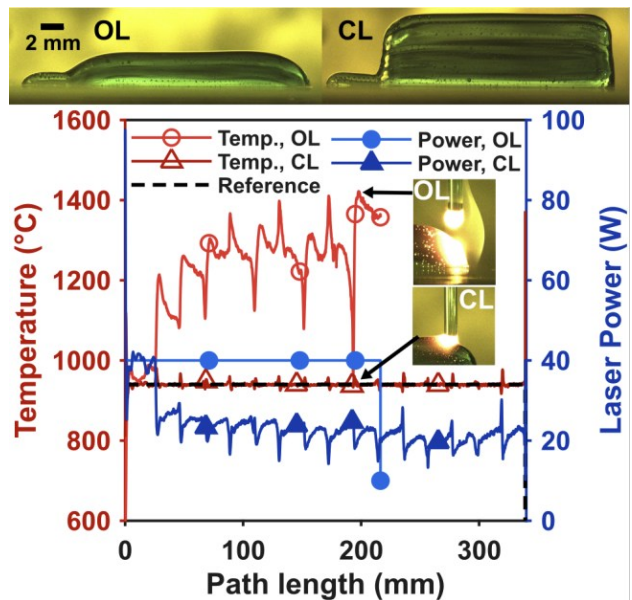

**Figure 13:** Spatial temperature and laser power trajectories for 16-layer walls 20 mm in length with $f = v$ = 0.5 mm/s, $d_l$ = 0.6 mm, and $d_f$ = 5 mm with open loop control ($L$ = 40 W) and closed loop control ($T_r$ = 940 °C) depositions. Images show prints at ninth corner, and final fabricated parts.

## 6. Summary and Conclusions

A real-time control system capable of synchronizing process parameter, thermal camera, and visual camera data in the DGF process was introduced in this paper. A process parameter map for $f = v = 0.5$ mm/s was constructed. A data-driven dynamic temperature process model was experimentally built and a closed loop temperature controller was designed. The closed loop controller is implemented during fabrication and utilizes the Internal Model Principle to reject constant and slowly varying disturbances. A series of experimental studies were conducted.

The process parameter map showed that track work zone temperature increased as the laser power increased and the distance above the focus decreased. The first-order, linear data-driven dynamic temperature process model showed > 70% agreement with validation data. When fabricating tracks with $d_f = 10$ mm and $L = 70$ W, the track periodically wetted to the substrate and then detached from the substrate, producing a failed build. By using the closed loop temperature controller with $T_r = 800$ °C, a complete track with a consistent morphology was fabricated. When fabricating tracks with $d_f = 3$ mm and $L = 10$ and 30 W, the track did not wet to the substrate and the track completely detached from the substrate, respectively. By using the closed loop temperature controller with $T_r = 850$ °C, a complete track with a consistent morphology was fabricated. During this experiment, the average laser power was 20 W, which was used in a subsequent constant laser power experiment. In this experiment, the track detached from the substrate half way through the experiment. Therefore, closed loop temperature control is able to extend the process parameter map in the DGF process.

When fabricating walls with $L = 40$ W, the first few layers were built well; however, the print failed at the ninth layer when the filament completely detached from the part. The failure is due to the fact that the heat transfer characteristics change as the wall is being built since the conduction path dramatically changes (i.e., from a large substrate that acts as a heat sink to a thin wall that acts as an insulator). Therefore, the process map developed for single tracks fabricated on a substrate is no longer applicable. Further, there was excessive heat at the corners that caused material vaporization and the edges to be excessively rounded. The closed loop controller with $T_r = 940$ °C is able to fabricate a wall with good morphology. The average laser power decreases each layer as the controller accounts for the change in the heat transfer characteristics. Also, the vaporization at the corners did not occur and the fabricated wall had straighter sides.

**CRediT authorship contribution statement**

**Balark Tiwari:** Conceptualization, Methodology, Software, Investigation, Resources, Formal analysis, Validation, Data curation, Writing – original draft, Writing – review and editing. **Nishan Khadka:** Writing – review and editing. **Nicholas Capps:** Writing – review and editing. **Andre Bos:** Supervision, Project administration. **Douglas Meredith:** Supervision, Project administration. **John Bernardin:** Project administration. **Edward C. Kinzel:** Writing – review and editing, Project administration, Supervision, Funding acquisition. **Robert Landers:** Conceptualization, Methodology, Validation, Project administration, Writing – review and editing, Supervision, Funding acquisition.

## Declaration of Generative AI and AI-assisted technologies in the manuscript preparation process

During the preparation of this work the authors used ChatGPT (OpenAI) in order to enhance sentence flow and correct grammar during manuscript preparation. After using this tool/service, the authors reviewed and edited the content as needed and take full responsibility for the content of the published article.

## Declaration of competing interest

The authors declare that they have no known competing financial interests or personal relationships that could have appeared to influence the work reported in this paper.

## 7. Acknowledgment

This research was supported by Los Alamos National Laboratory, under Contract CW7080.